\documentclass[useAMS,usenatbib]{mn2e}

\usepackage{graphicx}
\usepackage{amsmath}
\usepackage{natbib}
\usepackage{color}

\title[A generalized maximum volume density estimator]{A maximum volume density estimator generalized over a proper motion- limited sample}
\author[M. C. Lam, N. Rowell and N. C. Hambly]{Marco C. Lam\thanks{E-mail: mlam@roe.ac.uk}, Nicholas Rowell, Nigel C. Hambly\\Institute for Astronomy, University of Edinburgh, Royal Observatory of Edinburgh, Blackford Hill, Edinburgh EH9 3HJ, UK}

\begin{document}

\date{Accepted . Received ; in original form }

\pagerange{\pageref{firstpage}--\pageref{lastpage}} \pubyear{2015}

\maketitle

\label{firstpage}

\begin{abstract}
The traditional Schmidt density estimator has been proven to be unbiased and effective in a magnitude-limited sample. Previously, efforts have been made to generalize it for populations with non-uniform density and proper motion-limited cases. This work shows that the then good assumptions for a proper motion-limited sample are no longer sufficient to cope with modern data. Populations with larger differences in the kinematics as compared to the Local Standard of Rest are most severely affected. We show that this systematic bias can be removed by treating the discovery fraction inseparable from the generalized maximum volume integrand. The treatment can be applied to any proper motion-limited sample with good knowledge of the kinematics. This work demonstrates the method through application to a mock catalogue of a white dwarf-only solar neighbourhood for various scenarios and compared against the traditional treatment using a survey with Pan-STARRS-like characteristics.
\end{abstract}


\section{Introduction}
The use of a maximum volume as a density estimator began when \citet{1968ApJ...151..393S} introduced the V/V$_{max}$ technique for analysing the luminosity function of quasars, $\phi_{s}$, where V and V$_{max}$ are the volume enclosed by the object and the volume in which the object can be found at the given survey limits respectively and $\phi_{s}$ is the number density. The estimator can be used as a means of testing the completeness of a sample simultaneously. \citet{1976ApJ...207..700F} showed that $\phi_{s}$ is unbiased and superior to the older ``classical" estimator (N/V) when the magnitude or luminosity bins are not small. He further described the procedure of combining the density estimator from two non-overlapping areas.

Information from a single survey is limited due to the relatively small number of objects. When several complete samples are combined, more fundamental parameters of the population of objects can be determined, and with smaller associated uncertainties. In the light of this problem, \citet{1980ApJ...235..694A} investigated different ways of combining catalogues, namely the incoherent region-independent method, the incoherent domain-independent method and the coherent method. All of these are superior over the original Schmidt method, with the coherent method being most accurate. When the effects of a space-density
gradient were corrected for \citep{1989MNRAS.238..709S, 1993ApJ...414..254T, 2015PASP} this method was extended to estimate stellar density where the density profile of the Galaxy varies significantly along different lines of sight. Because of the small distances probed, only the scale height effects are considered while the scale length is assumed to be constant.

In order to consider a sample of proper motion objects, \citet{1975ApJ...202...22S} extended his estimator to cope with both photometric and proper motion detection limits. The new estimator considers the tangential velocity as an intrinsic property of an object such that it can be kept as a constant. Then, the distance limits can be found easily by applying the upper and lower proper motion limits of the survey to a simple relation between tangential velocity, proper motion and distance: $v_{tan} = 4.74\mu D$~km\,s$^{-1}$, where $\mu$ is the proper motion in arcseconds per year and $D$ is the distance to the object in parsecs.

Cool white dwarfs\,(WDs) and subdwarfs\,(sds) have similar optical colours to the main sequence stars\,(MSS) while those of brown dwarfs (BDs) are similar to the giants. Therefore, it is difficult to distinguish them in colour-colour space. Due to their small radii, WDs, sds and BDs are located far from MSSs and giants in the HR diagram. However, when objects are only detected in a few broadband filters, it is impossible to classify them reliably which would lead to poor object selections and distance estimates. To overcome this problem, it is common to use reduced proper motion\,(RPM) as a crude estimate of absolute magnitude to separate samples of subluminous objects from higher luminosity contaminants. In order to obtain a clean sample of WDs\,(for example an extreme subdwarf would easily be confused with a WDs with low tangential velocity) a lower tangential velocity has to be applied to remove the ambiguous objects. This procedure introduces an incompleteness which has to be corrected for. This problem was identified by \citet{1986ApJ...308..347B} and \citet{1992MNRAS.255..521E} separately. The former adopted a Monte Carlo\,(MC) simulation approach to correct for the incompleteness, while the latter was done analytically. However, the two methods evolved separately. In the simulation front, \citet{1999ASPC..169...51L} constructed some simulations based on different Galactic models to study the incompleteness due to proper motion selection after some strong arguments\,(\citealt{1995LNP...443...24O}, \citealt{1996Natur.382..692O}) pointing towards an incomplete LHS catalogue used in earlier studies\,(eg. \citealt{1987ApJ...315L..77W}, \citealt{1988ApJ...332..891L}). This correction, known as the discovery fraction, $\chi$, was then applied by \citeauthor{2006AJ....131..571H}\,(\citeyear{2006AJ....131..571H}, hereafter H06). On the other hand, \citet{1999MNRAS.306..736K} used the analytical approach to correct for the incompleteness. Instead of calculating the discovery fractions from integrating over the density profile, \citet{2003MNRAS.344..583D} arrived at the discovery fractions by integrating over the Schwarzschild distribution functions. \citeauthor{2011MNRAS.417...93R}\,(\citeyear{2011MNRAS.417...93R}, hereafter RH11) further generalized the technique to cope with an all sky survey as opposed to the individual fields of view employed in earlier works.

This work studies the discovery fraction in detail and shows that the discovery fraction has to be incorporated into the volume integral in order to arrive at a correct density estimation. In the next Section, we discuss the method of the simulation of the solar neighbourhood. In Section~3 we discuss different maximum volume estimators and discovery fractions and how their shortcomings can be removed by a new approach. This new method is then applied to a simulated sample of white dwarfs in Section~4, where we choose survey parameters typical of the state--of--the--art Pan--STARRS optical sky survey~(\citealt{2013ASPC..469..253H} and references therein). In the last Section, we discuss the possible extension of the method and conclude this work.

\section{Population Synthesis}
\label{sec:mcsimulation}
Monte Carlo simulations are used to produce snapshots of WD-only solar neighbourhoods which carry six dimensional phase space information. The volume probed in this work is assumed to be small such that the simulation is done in a Cartesian space, instead of a plane polar system centred at the Galactic Centre\,(GC). The Galaxy is further assumed to have three distinct kinematic components: a thin disc, a thick disc and a stellar halo, all of which have no density variations along the co-planar direction of the Galactic plane. All vertical structures follow exponential profiles, with scale height H. The velocity components, $U$, $V$ and $W$, of each object are drawn from the Gaussian distributions constructed from the measured means and standard deviations of the three sets of kinematics that describes the three populations in the solar neighbourhood. The thin and thick disc populations are assigned with constant star formation rates since look back time, $\tau = 8$\,Gyr and $\tau=10$\,Gyr  respectively while the halo has a star burst of duration 1\,Gyr at $\tau=12.5$\,Gyr. The initial mass function has an exponent of -2.3 \citep{2001MNRAS.322..231K}, and the initial-final mass function(IFMF) follows the ones in \citet{2009ApJ...705..408K}
\begin{align}
m_{f} = 
\begin{cases}
0.101m_{i} + 0.463, & 0.5M_{\odot}<m_{i}\leq4.0M_{\odot} \\
0.047m_{i} + 0.679, & 4.0M_{\odot}<m_{i}\leq7.0M_{\odot}.
\end{cases}
\end{align}
The MS life time has to be added in order to calculate the cooling time, and hence the magnitude of a WD. We have adopted the stellar evolution tracks from the Padova group (PARSEC;~\citealt{2012MNRAS.427..127B}) with a metallicity of $Z=0.019$ and $Y=0.30$~\citep{2000A&AS..141..371G}. Together with the pure hydrogen WD\,(DA) atmosphere cooling models with constant surface gravity $\log{g}=8.0$ and synthetic colours\footnote{http://www.astro.umontreal.ca/$\sim$bergeron/CoolingModels/} (\citealt{2006AJ....132.1221H}, \citealt{2006ApJ...651L.137K}, \citealt{2011ApJ...730..128T} and \citealt{2011ApJ...737...28B}), a theoretical luminosity function\,(LF) is produced which can then be used as the PDF in the Monte Carlo simulation. The normalisations of the PDFs are adopted from the WD densities found in RH11. The volume in which objects are distributed is limited to half a magnitude deeper than the maximum distance at which the survey can probe given its brightness. The half magnitude is to allow for random fluctuations near the detection limit after noise is added. The input parameters are assumed to be invariant with time and are summarized in Table~1.\\
\begin{table}
\centering
\caption{Physical properties of the Galaxy used in the Monte Carlo simulation.}
\begin{tabular}{ l *{3}{c} }
\hline
\hline
Parameter & Thin disc & Thick disc & Stellar Halo\\
\hline
$\langle$U$\rangle$/km s$^{-1}$ & -8.62$^{a}$ & -11.0$^{d}$ & -26.0$^{d}$ \\
$\langle$V$\rangle$/km s$^{-1}$ & -20.04$^{a}$ & -42.0$^{d}$ & -199.0$^{d}$ \\
$\langle$W$\rangle$/km s$^{-1}$ & -7.10$^{a}$ & -12.0$^{d}$ & -12.0$^{d}$ \\
$\sigma_{U}$/km s$^{-1}$ & 32.4$^{a}$ & 50.0$^{d}$ & 141.0$^{d}$ \\
$\sigma_{V}$/km s$^{-1}$ & 23.0$^{a}$ & 56.0$^{d}$ & 106.0$^{d}$ \\
$\sigma_{W}$/km s$^{-1}$ & 18.1$^{a}$ & 34.0$^{d}$ & 94.0$^{d}$ \\
H/pc & 250$^{b}$ & 780$^{e}$ & $\infty$ \\
n/pc$^{-3}$ & 0.00310$^{c}$& 0.00064$^{c}$& 0.00019$^{c}$ \\
\hline
\end{tabular}
\begin{enumerate}
\itemsep0em
\item[a] \citealt{2009AJ....137..266F}
\item[b] \citealt{1998A&A...333..106M}
\item[c] RH11
\item[d] \citealt{2000AJ....119.2843C}
\item[e] \citealt{2006AJ....132.1768G}
\end{enumerate}
\end{table}
\indent
From the true distance and true bolometric magnitude drawn from the PDF, the true apparent magnitudes in the Pan--STARRS g$_{\rm P1}$, r$_{\rm P1}$, i$_{\rm P1}$, z$_{\rm P1}$ and y$_{\rm P1}$ filters are assigned \citep{2012ApJ...750...99T, 2012ApJ...756..158S, 2013ApJS..205...20M}. The uncertainties in those filters, $\sigma_{m_{i}}$, are assumed to scale exponentially with magnitude and  are described by
\begin{equation}
\label{eq:uncertainties}
\sigma_{m_{i}} = a_{i} \times \exp^{(m_{i}-15.0)} + b_{i}
\end{equation}
where $a_{i}$ and $b_{i}$ are constants measured from the first Pan--STARRS survey~(PS1; magnitude subscript `P1') at Processing Version~(PV)~1.1 and $m_{i}$ is the magnitude in filter $i$ (Fig. 1). Realistic dispersion is added to the uncertainties by resampling $\sigma_{m_{i}}$ with a Gaussian distribution with standard deviations of $0.1\times\sigma_{m_{i}}$ centred at the noiseless $\sigma_{m_{i}}$. The magnitudes in each filter are then drawn from a Gaussian distribution with a standard deviation $\sigma_{m_{i}}$. The proper motion uncertainty is based on the r$_{\rm P1}$ magnitude.
\begin{table}
\centering
\caption{Parameters for the noise model.}
\begin{tabular}{ *{3}{c} }
\hline
\hline
Filter or Proper Motion & $a_{i}/mag$ & $b_{i}/mag$ \\
\hline
g         & 0.000125 & 0.00065 \\
r          & 0.000120 & 0.00075 \\
i          & 0.000175 & 0.00065 \\
z         & 0.000325 & 0.00089 \\
y         & 0.000750 & 0.00120 \\
$\mu$ & 0.000300 & 0.00050 \\
\hline
\end{tabular}
\end{table}

\begin{figure}
\begin{center}
\includegraphics[width=84mm]{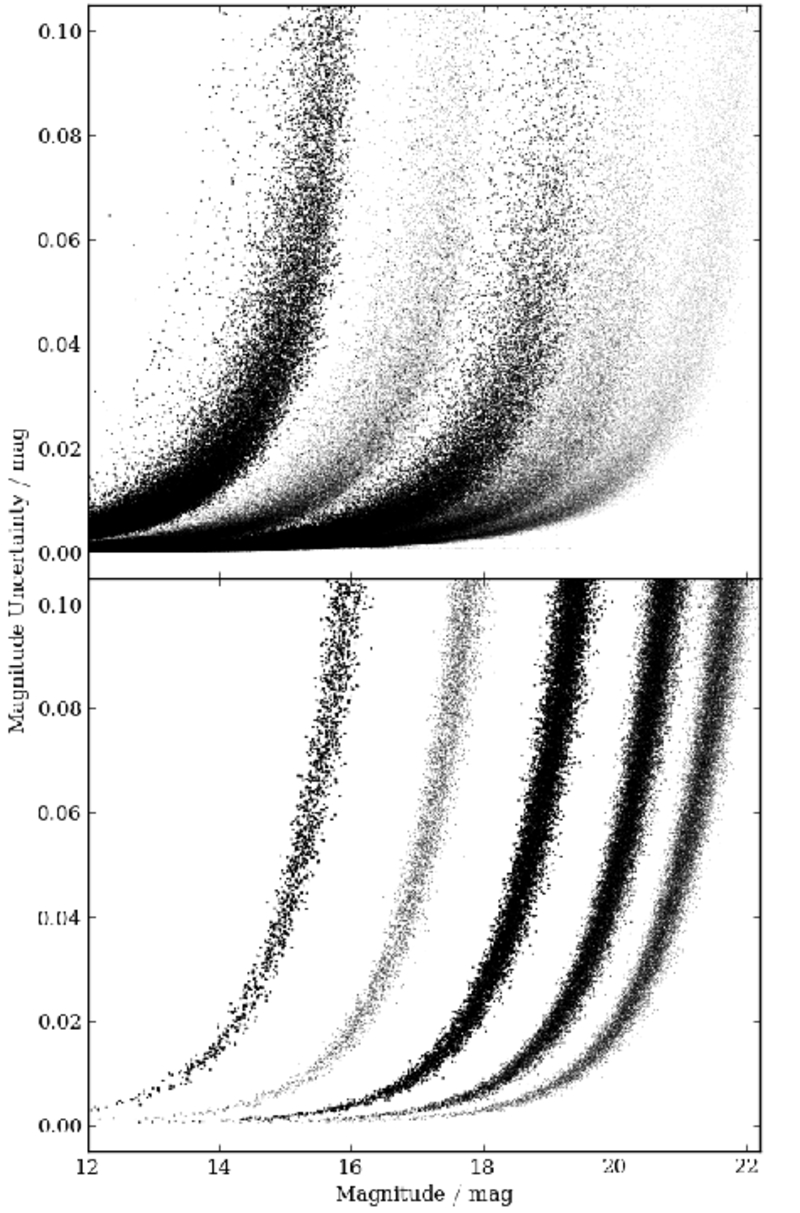}
\caption{Top: the magnitudes and the associated uncertainties of point sources with 5$\sigma$ proper motions in the PV1.1 of Pan--STARRS, from right to left g$_{\rm P1}$, r$_{\rm P1}$, i$_{\rm P1}$, z$_{\rm P1}$ and y$_{\rm P1}$ filters. Each successive filter is offset by one magnitude. Bottom: the magnitudes and uncertainties distribution of WDs reproduced in the simulation.}
\end{center}
\end{figure}

\section{generalized Maximum Volume Density Estimator}
The classical estimator $\Phi = N/$V for a volume-limited sample is of little practical use for analysing small numbers of objects or strongly localized and kinematically biased groups of stars that are selected by proper motion and apparent magnitude. For these samples, the 1/V$_{max}$ method is generally regarded as a superior estimator of the LF \citep{1976ApJ...207..700F}. The contribution of each object to the LF is weighted by the inverse of the maximum volume in which an object could be observed by the survey. For example, for a given bin of objects with index k, the space density is the sum of all 1/V$_{max}$
\begin{equation}
\Phi_{k} = \displaystyle\sum\limits_{i=1}^{N_{k}} \frac{1}{\text{V}_{max, i}}
\end{equation}
for $N_{k}$ objects in the k$^{th}$ bin. The uncertainty of each star's contribution is assumed to follow Poisson statistics. The sum of all errors in quadrature within a luminosity bin is therefore,
\begin{equation}
\sigma_{k} = \left[\displaystyle\sum\limits_{i=1}^{N_{k}} \left(\frac{1}{\text{V}_{max, i}}\right)^{2}\right]^{\frac{1}{2}}.
\end{equation}
The traditional 1/V$_{max}$ technique assumes that objects are uniformly distributed in space. However, in reality, stars in the solar neighbourhood are concentrated in the plane of the disc. The effects of space-density gradient can be corrected by assuming a density law and defining a maximum generalized volume V$_{gen}$ (\citealp{1989MNRAS.238..709S}; \citealp{1993ApJ...414..254T}) which is calculated by integrating the appropriate stellar density profile $\rho/\rho_{\odot}$ along the line of sight between the minimum distance, $d_{min}$, and maximum distance, $d_{max}$. This leads to the integral
\begin{equation}
\text{V}_{gen,S89} = \Omega \int^{d_{max}}_{d_{min}} \frac{\rho(r)}{\rho_{\odot}} \, r^{2}\,dr
\end{equation}
where $\Omega$ is the size of the solid angle of the survey. To minimise the contamination from extreme subdwarfs scattered into the WD regime in RPM-colour space, a lower tangential velocity limit, $v_{tan,lower}$, is applied to remove most of the contaminants. Traditionally, the discovery fraction, $\chi_{v}$ which is the fraction of objects with tangential velocities larger than the lower tangential velocity limit, is only Galactic model and survey-footprint dependent (\citealp{1986ApJ...308..347B}, \citealp{1999ASPC..169...51L}, \citealp{2006AJ....131..571H}) such that the maximum volume density estimator can be written in the form
\begin{equation}
\text{V}_{gen,H06} = \chi_{v}(v_{tan,lower}) \, \Omega \int^{d_{max}}_{d_{min}} \frac{\rho(r)}{\rho_{\odot}} \, r^{2} \, dr
\end{equation}
where the distance limits are derived from both photometric and proper motion limits of the survey by calculating
\begin{equation}
\label{eq:dmin}
d_{min} = d \times \text{max} \left[ 10^{\frac{(m_{min,i}-M_{i})}{5}}, \frac{\mu}{\mu_{max}} \right]
\end{equation}
\begin{equation}
\label{eq:dmax}
d_{max} = d \times \text{min} \left[ 10^{\frac{(m_{max,i}-M_{i})}{5}}, \frac{\mu}{\mu_{min}} \right]
\end{equation}
where $m_{min,i}$, $m_{max,i}$ and $M_{i}$ are the photometric limits and the absolute magnitudes of the object in filter $i$ respectively. The proper motion terms, $\mu/\mu_{max}$ and $\mu/\mu_{min}$, are rationalized by assuming an object would carry the same tangential velocity if it were placed closer to or farther from the observer (analogous to the absolute magnitude) and/or in an arbitrary line of sight.

\subsection{Attempt to Modify the Discovery Fraction}
RH11 extended the $\chi_{v}$ to include a directional dependence in order to account for the varying survey properties and stellar tangential velocity distribution across the sky. In RH11, the Schwarzschild distribution function is used instead to calculate the tangential velocity distribution, $P(v_{tan})$, analytically. The discovery fraction can be found by projecting the velocity ellipsoid onto the tangent plane of observation \citep{1983veas.book.....M}. Thus it allows one to arrive at a precise $\chi(\alpha,\delta)$ without taking the average properties over a large area. Therefore, the volume integral can be modified to
\begin{equation}
\text{V}_{gen,RH11} = \sum_{i} \Omega_{i} \, \chi_{v}(i,v_{tan,lower}) \,  \int^{d_{max}}_{d_{min}} \frac{\rho_{i}(r)}{\rho_{\odot}} \, r^{2} \, dr
\end{equation}
where $i$ denotes each sky cell covered by a Schmidt survey field employed in the production of the catalogue used by RH11 (hereafter the generalized method). This should, in theory, have taken into account all the small scale variations which a positional-independent $\chi_{v}$ would not be able to deal with.
\subsection{Problems in this framework}
We have identified a bias present in the {\it whole family} of maximum volume methods when calculating distance limits by holding the tangential velocity constant along a single line of sight \citep{1975ApJ...202...22S}; and the consequence to the the discovery fraction applied in H06 and RH11 when the tangential velocity is further held constant across the sky. These can be described as follows:
\subsection*{Constant tangential velocity along line of sight}
The kinematics of an object is a property of the Galaxy. An object at a given magnitude at any given distance from the observer should not carry the same tangential velocity at a different line of sight distance when tested for observability. This assumption is only good over small field of views and small range of line of sight distances, while an acceptable size of the smallness is very difficult to access if not unquantifiable.
\subsection*{Constant tangential velocity across the sky}
Consider a spatially uniform population like the stellar halo, and an all-sky survey where the proper motion limits are the same over the whole sky. The tangential velocity distribution varies along different lines of sight due to the solar motion. For stars at a given magnitude, a different fraction of the population will pass the proper motion limits along different lines of sight due to the differences in the tangential velocity distributions. In the most extreme cases, on average, a halo WD observed in the direction of the Anti-Galactic Center~(AGC) would appear to have a large velocity due to reflex motion imparted by the Sun in its orbit within the Galaxy. However, if one is observed in the direction to the solar apex instead, the motion would be much smaller on average. The consequence is that when proper motion limits and tangential velocity limits are applied, different numbers of stars would be detected in different regions of sky even for identical survey limits and spatial density. That is the motivation for RH11 to consider different tangential velocity distributions along different lines of sight. The method would be completely correct in the framework where {\it the tangential velocity is an intrinsic property of an object}. However, in the case where tangential velocity is not constant, there would be a mismatch between the parameter space which the discovery fraction and the maximum volume explore~(see Section~\ref{sec:new_approach} where we will demonstrate how the new approach can solve both problems).

\subsection{A Closer Look at the Discovery Fraction}
\label{sec:closer_look}
\begin{figure}
\begin{center}
\includegraphics[width=84mm]{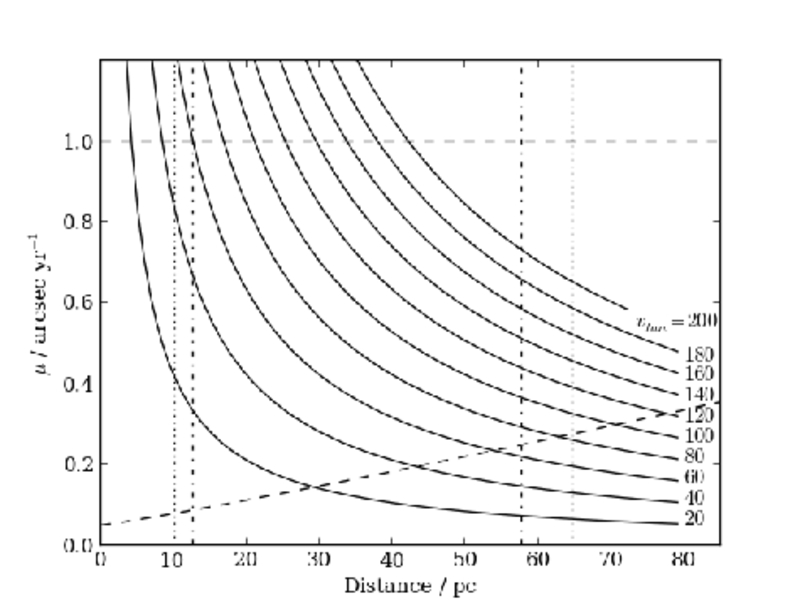}
\end{center}
\caption{
This plot illustrates how tangential velocity and proper motion limits behave in the $\mu-D$ space. The solid lines are the contours of the tangential velocities from 20 to 200\,km\,s$^{-1}$ in steps of 20\,km\,s$^{-1}$. The photometric distance limits mark the range of distances in which an object can be placed and stay within the detection limits. The dot-dashed lines are the distance limits calculated from Equations \ref{eq:dmin} and \ref{eq:dmax}, which is by fixing the tangential velocity of an object such that a proper motion limit correspond to a fixed distance limit. The dotted lines are the distance limits calculated from Equations \ref{eq:dmin_phot} and \ref{eq:dmax_phot}, where the distance limits are not functions of the kinematics. The dashed lines are the upper and lower proper motion limits; in the latter case, this limit rises as astrometric errors rise for increasingly faint flux levels as distance increases.}
\label{fig:vtan_mu}
\end{figure}

\begin{figure}
\begin{center}
\includegraphics[width=84mm]{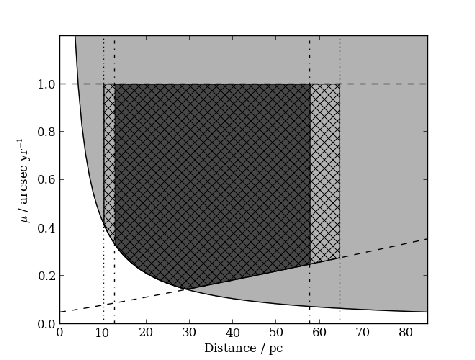}
\caption{
The area in light grey~(which also includes the hatched and dark grey areas) is the parameter space where the discovery fraction is calculated when the proper motion limits and distance limits are not considered~(e.g. H06 and RH11). The dark grey area corresponds to the generalized volume where the distance limits~(Eq. \ref{eq:dmin} and \ref{eq:dmax}) are considered, it is clear that there are inconsistencies between the two parameter space. The hatched area is the parameter space where both the discovery fraction and the volume are integrated over in the modified method~(Please note that in this case the distance limits are found from Eq.~\ref{eq:dmin_phot} and \ref{eq:dmax_phot}). The weighted area gives the discovery fraction in the new method~(see Figure \ref{fig:discovery_fraction} for the weight maps). A treatment without considering the effects of the proper motion limits and distance limits on the parameter space in which the discovery fraction is integrated over would always overestimates the completeness, which translates to an underestimation in the number density.}
\label{fig:different_discovery_fraction}
\end{center}
\end{figure}

\begin{figure}
\begin{center}
\includegraphics[width=84mm]{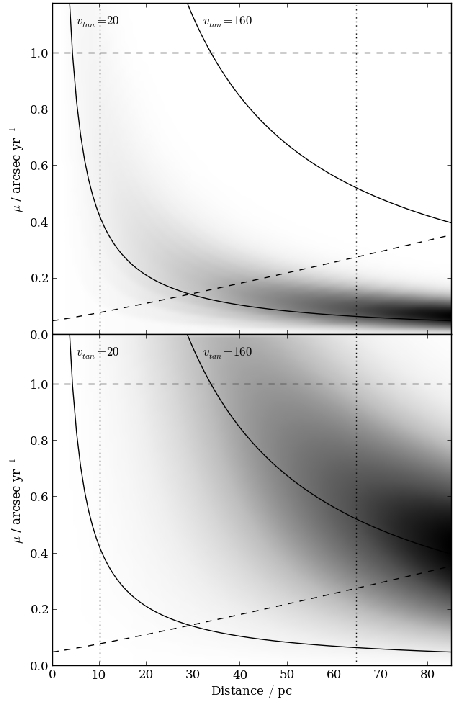}
\end{center}
\caption{
The top panel shows the probability distribution as a function of proper motion and distance of a volume limited thin disc population in the direction of $(\alpha,\delta)=(0.0,0.0)$ over-plotted with the proper motions, tangential velocities and distance limits. The discovery fraction is the weighted area within the specified limits where the probability distribution function is used as the weights. The lower panel shows those of a halo population. The heaviest part of the thin disc weight function is excluded by the proper motion limit, but much more of the heavily weighted region of the stellar halo is within the allowed parameter space, so the tangential velocity limit at $\sim20$\,km\,s$^{-1}$ has a much smaller effect in the thin disc than in the halo.}
\label{fig:discovery_fraction}
\end{figure}

Proper motions, tangential velocities and distances are related by a simple equation
\begin{equation}
\label{equation:vtan_mu_d}
v_{tan} \approx 4.74\,\mu\,D.
\end{equation}
All the survey limits can be shown in the proper motion-distance space~(Fig.~\ref{fig:vtan_mu}). A valid approach should not impose any assumptions that restrict an object in this two dimensional space. The discovery fraction is equal to the weighted area restricted by the survey limits, where the weight map~(Fig.~\ref{fig:discovery_fraction}) is generated from dividing the tangential velocity distribution by the distances,
\begin{equation}
W(\mu,D)=\frac{P(v_{tan})}{4.74 \times D}.
\end{equation}
In the generalized method, $v_{tan}$ is fixed and is thus separable from $\mu$, which follow the lines of constant tangential velocity at $v_{tan}=$ 20 - 200 km\,s$^{-1}$ in 20\,km\,s$^{-1}$ interval in Fig.\,\ref{fig:vtan_mu}. However, the two variables are related through distance as shown in Equation\,\ref{equation:vtan_mu_d}. This implies a selection criterion in one parameter would lead to a selection effect in the other. The generalized method only deals with the tangential velocity limits but ignores the consequential effect of the change in the proper motion limits bounding the discovery fractions. A similar effect also appears in the treatment of the distance limits. The volume integral is bounded by the maximum and minimum distances in which an object can be found but the discovery fraction in RH11 includes everything above the line $v_{tan}=20$\,km\,s$^{-1}$ as represented by the light grey area in Fig.\,\ref{fig:different_discovery_fraction}. The true discovery fraction should be bounded by the same limits as applied to the volume integral (hatched area). Thus, $\chi_{v}$ is always overestimated which translates to an underestimation in the luminosity function. This effect is stronger for:
\begin{itemize}
\item[i)] a survey with small upper proper motion limit because the lower this limit, the larger the overestimation of the discovery fraction. This can be seen from Fig. \ref{fig:different_discovery_fraction} where a smaller upper proper motion limit would lead to a smaller cross-hatched area, which means the discovery fraction would be overestimated.
\item[ii)] a population with large differences in the kinematics compared to the observer because objects tend to have large proper motions. This shifts the region with the highest probability in $W(\mu,D)$ (Fig.~\ref{fig:discovery_fraction}) to larger proper motions. In the case of the Galaxy, the stellar halo is the most susceptible to this effect.
\item[iii)] intrinsically faint objects that carry small maximum observable distance. It is clear from Fig.\,\ref{fig:discovery_fraction} that the relative $W(\mu,D)$ at low proper motion increases with distance when the $P(v_{tan})/D$ peaks at smaller proper motion as distance increases. This affects the discovery fraction more severely when the lower proper motion limit is large since more heavy-weighted area would be included in, for example, both the H06 and RH11 methods.
\end{itemize}
\indent
As an example of a current survey which can be employed in WDLF studies \citep{2013ASPC..469..253H}, the Pan--STARRS upper proper motion limit is O$(1)$"\,yr$^{-1}$ which means the first problem may affect the analysis of WDLF without proper treatment. Halo WDs have higher velocities and are older~(i.e.~fainter) than those of the discs and hence the effect on the halo population would be much larger than that in the discs. As it is of great interest to probe the halo WDLF at such low luminosity to explore the possible scenarios of the star formation history of the Galaxy, it is necessary to correct this bias.

\subsection{A New Approach}
\label{sec:new_approach}
In order to compute the discovery fraction properly, the parameter space in which an object could be observed by the survey has to be identical to that used in the discovery fraction integral. For each step of the numerical integration, it is necessary to calculate the instantaneous discovery fraction which is limited by the upper and lower proper motion limits, as well as the tangential velocity limits. It is worth mentioning that in the effective volume method in RH11 a similar approach was adopted that would have corrected for the bias noted in the last Section although the bias was not explicitly identified and discussed in that work. Instead of dealing in an object by object basis, their correction was applied statistically. The strength of that method is that the WDLFs of the three components could be untangled. However, binning objects by their bolometric magnitudes before membership association would lead to a loss of information. Furthermore, it loses the generality so that it cannot be applied to other luminosity estimators. In order to keep the analysis in an object by object basis, one should consider the {\it modified volume} integral. It is computed by integrating the physical survey volume along the line of sight, and considering at each distance step both the stellar density profile and the fraction of objects that pass the tangential velocity limits implied by the proper motion limits. The total generalized survey volume between d$_{\text{min}}$ and d$_{\text{max}}$ is therefore calculated by
\begin{equation}
\label{eq:volume_new}
V_{mod} = \Omega \int_{d_{\text{min}}}^{d_{\text{max}}} \frac{\rho(r)}{\rho_{\odot}} r^2 \left[ \int_{a(r)}^{b(r)} P(v_{\mathrm{T}}) dv_{\mathrm{T}} \right] dr,
\end{equation}
where the distance limits are solely determined by the photometric limits of the survey
\begin{equation}
\label{eq:dmin_phot}
d_{min} = d \times max \left[ 10^{\frac{(m_{min,i}-M_{i})}{5}} \right]
\end{equation}
\begin{equation}
\label{eq:dmax_phot}
d_{max} = d \times min \left[ 10^{\frac{(m_{max,i}-M_{i})}{5}} \right] ,
\end{equation}
$P(v_{\mathrm{T}})$ is the tangential velocity distribution, $\rho(r)$ corrects for a non-uniform population density profile (i.e. disc populations), and $\Omega$ is the survey footprint area in steradians. Note that $P(v_{\mathrm{T}})$ and $\rho(r)$ depend on the line of sight, so this model only holds for small fields. The limits on the tangential velocity integral are
\begin{align}
a(r)& = \text{max}(v_{\text{min}}, 4.74\mu_{\text{min}}r)\\
b(r)& = \text{min}(v_{\text{max}}, 4.74\mu_{\text{max}}r),
\end{align}
where $v_{\text{min}}$ and $v_{\text{max}}$ are any fixed tangential velocity limits applied to reduce contamination from other stellar populations, and $4.74\mu_{\text{min}}r$ and $4.74\mu_{\text{max}}r$ are the tangential velocity limits at distance $r$ arising from the proper motion limits. The appropriate limits on the integral are found by considering both of these effects.

The new volume integral has the distance limits decoupled from the kinematics, which are completely absorbed into the discovery fraction. The decoupling simultaneously means that regardless of how the kinematic behaviour changes with respect to the direction of observation, the entire $\mu$-$D$-$v_{tan}$ parameter space in which an object can be found is explored. When the discovery fraction has a dependence on distance, it cannot be separated from the integrand. This means that the discovery fraction varies from object to object in any given direction, so it is sufficiently general to take a more realistic form of velocity distribution. This in turn allows for a distance dependent tangential velocity distribution that describes the Galaxy more realistically.

\subsection*{Survey volume generalized for kinematic selection}
It is instructive to examine how the survey volume as a function of distance is changed when the full effects of kinematic selection are taken into account, i.e. considering that proper motion limits result in implicit tangential velocity limits that vary as a function of distance. The generalized 1/V$_{\text{max}}$ method does not consider this, and computes a constant discovery fraction only from any fixed, external tangential velocity limits that are applied e.g. to reduce contamination from certain types of object.

\begin{figure}
\begin{center}
\includegraphics[width=84mm]{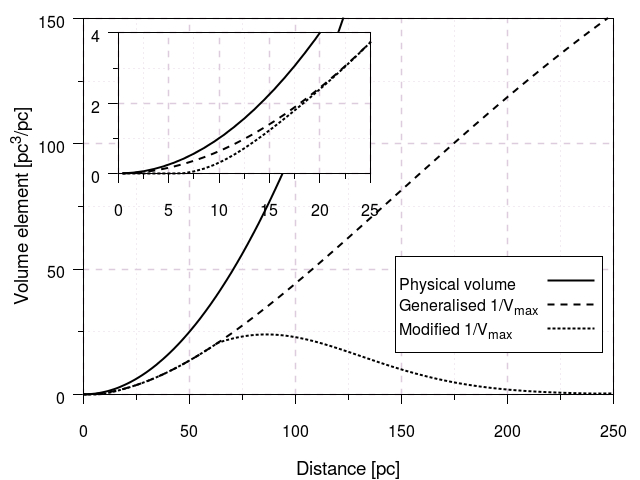}
\includegraphics[width=84mm]{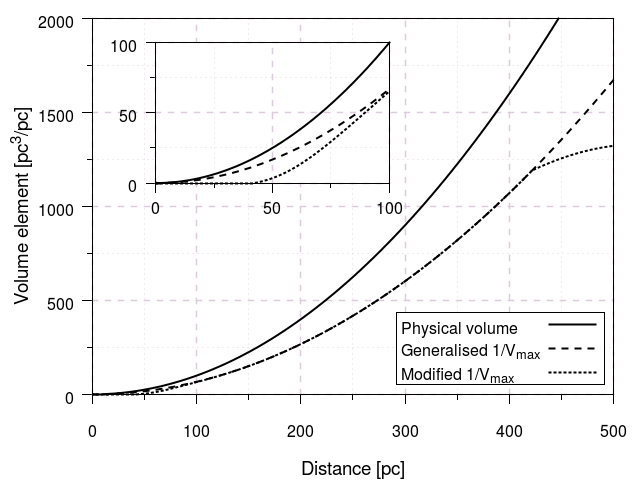}
\end{center}
\caption{generalized survey volume as a function of line of sight distance, for the stellar halo~(top) and thin disc~(bottom). The lines represent the true physical volume~(solid), the generalized volume~(dashed), and the modified volume~(dotted). The functions diverge at large distances where 4.74$\mu_{\text{min}}$r $>$ $v_{\text{min}}$~(see text) and the discovery fraction used by the generalized 1/V$_{\text{max}}$ method breaks down. A similar effect operates at small distances~(inset), where the upper proper motion limit results in a small upper tangential velocity limit that excludes a large fraction of the population.}
\label{fig:differential_volume}
\end{figure}

The differential survey volume as a function of distance, for the stellar halo and the thin disc, is presented in Fig.~\ref{fig:differential_volume}. These plots are computed for a line of sight towards the NGP, in order to exaggerate the effect of the non-uniform population density profile in the disc case, which falls off most rapidly in directions perpendicular to the plane. Halo and thin disc kinematics follow those in Table 1, with lower/upper proper motion limits of 0.1/1.0\,"\,yr$^{-1}$ and a survey solid angle of 0.01\,sr.

In the halo case, a fixed lower tangential velocity limit of $v_{\text{min}}$=200\,km\,s$^{-1}$ has been applied, which is the usual way of reducing contamination from disc WDs. $v_{\text{max}}$ has been left unconstrained; this corresponds to a constant discovery fraction of $\sim$0.67, which is the value used in the generalized 1/V$_{\text{max}}$ method leading to a generalized volume that is a constant fraction of the physical volume.

In the disc case, a fixed lower limit of $v_{\text{min}}$=30\,km\,s$^{-1}$ has been applied, which is used to reduce contamination from high velocity subdwarfs. $v_{\text{max}}$ is again unconstrained, leading to a constant discovery fraction of $\sim$0.66 used by the generalized 1/V$_{\text{max}}$ method, although note that in the disc case the generalized volume diverges from the true physical volume due to the non-uniform population density profile.

In both cases, there is a range of intermediate distances over which the generalized and modified 1/V$_{\text{max}}$ methods give the same result for the survey volume. This corresponds to the range over which the fixed external tangential velocity limits are active. Significant differences between the two methods arise at large distances, where the lower tangential velocity limit implied by the lower proper motion limit exceeds the fixed external limit. It is at these distances that the generalized 1/V$_{\text{max}}$ method, which fails to consider this effect, overestimates the generalized survey volume and thus underestimates the number density. A similar effect arises at small distances~(inset), where the upper tangential velocity limit implied by the upper proper motion limit is greatly reduced, causing most of the population to be excluded. The generalized 1/V$_{\text{max}}$ method, which considers only the fixed threshold $v_{\text{max}}$, fails to account for this effect and again overestimates the survey volume.

\section{Application to White Dwarf Luminosity Functions}
We have used the simulation outlined in Section \ref{sec:mcsimulation} to generate mock Pan--STARRS catalogues to compare the differences in applying the RH11 generalized volume and the new modified volume described in this work. The bright limits in all filters are set at 15.00\,mag, while the faint limits in g$_{\rm P1}$, r$_{\rm P1}$, i$_{\rm P1}$, z$_{\rm P1}$ and y$_{\rm P1}$ filters are at 21.50\,mag, 21.00\,mag, 20.50\,mag, 20.00\,mag and 20.00\,mag respectively. The lower proper motion limit is defined as five times the proper motion uncertainty at the given magnitude as defined in Equation \ref{eq:uncertainties}. The upper proper motion limit is set at 0.3"\,yr$^{-1}$ unless specified otherwise. Photometric parallaxes are found by fitting the magnitudes to the WD synthetic atmosphere models at fixed surface gravity $\log{g}=8.0$. Tests have shown that there exists local minima for effective temperature below 6000K so a Markov-Chain Monte Carlo method is used for minimisation. The implementation adopted the Python programme emcee\footnote{http://dan.iel.fm/emcee/} \citep{2013PASP..125..306F}. A lower tangential velocity limit is set at 20\,km\,s$^{-1}$.

\begin{figure}
\begin{center}
\includegraphics[width=84mm]{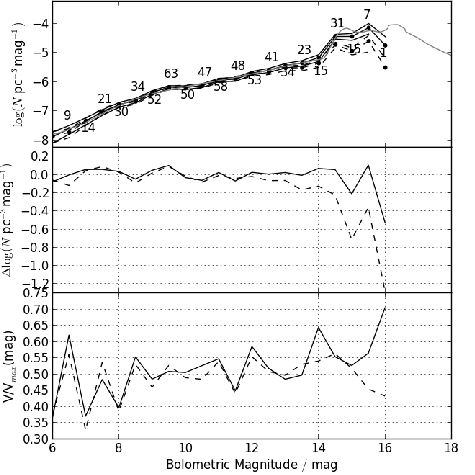}
\caption{
Top: WDLFs for a population with halo kinematics with the measured halo WD density. The two WDLFs are derived from the new approach~(solid) and the RH11 method~(dashed) overplotted with the input true WDLF~(grey). The lines above and below the WDLFs are the 1-sigma upper and lower limits. Middle: the deviations of the WDLF as a function of magnitude. The RH11 solution departs from the true LF at $\sim$12.0\,mag while the new approach at $\sim$15.0\,mag. Bottom: the $\langle$V/V$_{max} \rangle$ as a function of magnitude from the two methods. The uncertainties in $\langle$V/V$_{max} \rangle$ is $\frac{1}{\sqrt{12 N}}$, so in the $\langle$V/V$_{max} \rangle$s are within the statistical fluctuations up to 16\,mag.}
\label{fig:wdlf_new_old}
\end{center}
\end{figure}

\subsection{Stellar halo}
As discussed in Section~\ref{sec:closer_look}, faint objects and objects from a population with large difference in the kinematics from the observers are most susceptible to underestimation of the number density when using the generalized volume technique. Thus, the largest systematic errors are expected to be found in the faint end of the stellar halo WDLF. In the top panel of Fig.\ref{fig:wdlf_new_old}, the differences in the WDLF constructed by the two methods are shown. The two LFs agree with each other up to M$_{bol}\sim$12.0\,mag. Beyond that, the generalized method consistently underestimates the number density and the deviation increases as the objects get fainter. The maximum difference is more than 1.0\,dex. The Pan--STARRS filter $g_{\rm P1}$ reaches $\sim$400\,nm at the blue edge, so it is expected that the photometric parallax solutions become unreliable at the bright end, while with small numbers of objects in the faint end the statistical noise is significant. Most of the $\langle$V/V$_{max}\rangle$s in both cases are within 1$\sigma$ from 0.5, which is a necessary condition for an unbiased sample.

\subsection{Stellar Halo with 100 times the Observed Density}
\label{sec:stellar_halo_100}
\begin{figure}
\begin{center}
\includegraphics[width=84mm]{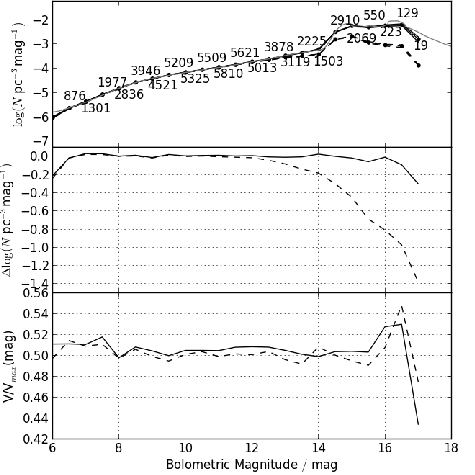}
\caption{Same as Fig.\ref{fig:wdlf_new_old} except the WD density is artificially inflated by 100 times to reduce random noise. With the increase number density, the WDLF departs a magnitude fainter in both cases. The $\langle$V/V$_{max} \rangle$ of the old method is still fluctuating about 0.5 even when the WDLF departs further and further away from the true LF. In the new method, the $\langle$V/V$_{max} \rangle$ is consistent within statistical uncertainties.}
\label{fig:wdlf_100_new_old}
\end{center}
\end{figure}

The stellar halo has a very low number density and it is well known that maximum volume estimators are prone to systematic bias with small number statistics. Thus, we have generated a stellar halo with 100 times the observed density to reduce such uncertainties. The WDLFs produced are more easily compared when the systematic uncertainties are much smaller and the corresponding plots are shown in Fig.~\ref{fig:wdlf_100_new_old}. The two LFs agree with each other up to M$_{bol}\sim12.0$\,mag again. With 100 times the density, faint objects down to 17.0\,mag (T$_{\text{eff}} \sim 3000$\,K) can be generated easily. The LF produced with the new method agrees with the input LF to luminosities down to $\sim$16.0\, while the RH11 method consistently underestimates the density, as predicted in Section \ref{sec:closer_look}. The $\langle$V/V$_{max}\rangle$ shows an interesting behaviour - the distribution is fluctuating about 0.5 regardless of the bias in the WDLF. It implies that $\langle$V/V$_{max}\rangle \approx 0.5$ is a necessary but not sufficient condition for both bias and completeness.

\subsection{Different upper proper motion limits}
\begin{figure}
\begin{center}
\includegraphics[width=84mm]{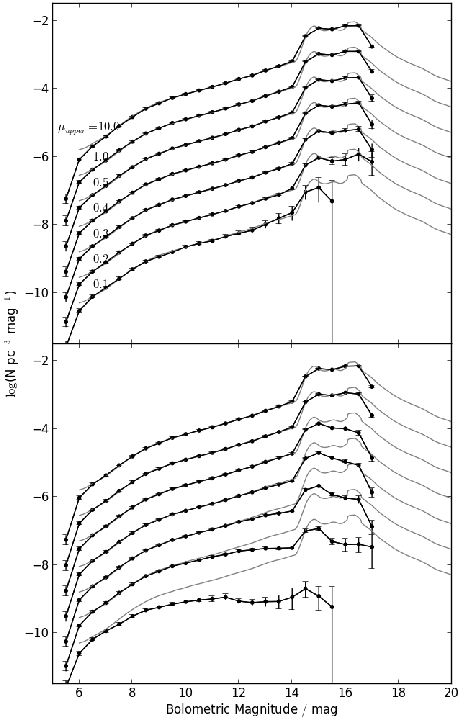}
\caption{Top: WFLD constructed with different upper proper motion limits at 100 times the measured density with the new modified volume method. From top to bottom, the upper proper motion limit is set at 10.0, 1.0, 0.5, 0.4, 0.3, 0.2 and 0.1"\,yr$^{-1}$. Bottom: same procedure repeated with the generalized volume. In the two sets with the smallest limits, dashed lines are used to join up the LFs at the faintest bin where the area under the dashed line should not be included in determining the total number density.}
\label{fig:different_upper_limits}
\end{center}
\end{figure}

As mentioned in Section~\ref{sec:closer_look}, a small upper proper motion limit would lead to a strong bias. This is illustrated in Fig\,\ref{fig:different_upper_limits}: as the upper proper motion limit increases, the old and new approaches converge. However, when the limit decreases, the underestimation in the number density increases when adopting the old method. From the bottom panel, even at an upper limit of 0.5"\,yr$^{-1}$, the generalized method fails to recover the WDLF beyond $\sim$15\,mag. With the new method, the number density estimation is recovered under any restricted proper motion selection within the statistical noise. In the modern and future surveys with the rapid photometric systems, the upper proper motion limit would be in the order of arcseconds so this effect would be small. However, when the pairing of the bright objects includes older photographic plate data, the upper proper motion limit would be severely hampered due to a large maximum epoch difference (e.g.~\citealp{2003MNRAS.344..583D} has an upper limit of 0.18"\,yr$^{-1}$ from pairing between POSS-I, POSS-II and SDSS). Another case is when samples at different velocity ranges are analyzed separately (e.g.~the effective volume method in RH11 where the discovery fraction was treated correctly) the restricted velocity range would amplify the shortcoming of the old method.

\subsection{Thin Disc, Thick disc and Stellar Halo}
\label{sec:galaxy}
\begin{figure}
\begin{center}
\includegraphics[width=84mm]{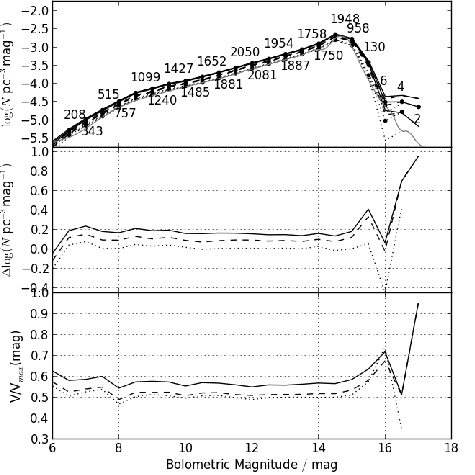}
\caption{
Top: WDLFs of the Galaxy constructed by assuming negligible halo contamination~(solid), with an additional fixed upper tangential velocity limit at 80\,km\,s$^{-1}$~(dashed) and by picking up only objects tagged as ``thin disc object" from the simulation~(dotted). Middle: the differences in the WDLF from the true LF as a function of magnitude. The overdensities are expected from the wrong association of kinematics model with the objects. Bottom: the $\langle$V/V$_{max}\rangle$ distributions as a function of magnitude. $\langle$V/V$_{max}\rangle$s are consistently larger than 0.5 implying objects tend to be found at larger distances than on average. This is expected from the objects with a uniform spatial density being associated with ones that have an exponential profile. The larger discovery fraction of the thin disc with respect to the halo amplifies the the effect further~(see text for detailed explanation).}
\label{fig:galaxy_upper_vtan}
\end{center}
\end{figure}

In the study of the Galactic WDLFs it is very difficult to assign population membership to individual objects. The consensus is that thin disc objects dominate the solar neighbourhood so that when studying the LF for the thin disc it is possible to assume a thin disc characteristic for all objects. This does not happen to be a good assumption as shown in the top panel of Fig. \ref{fig:galaxy_upper_vtan} where for the purposes of this exercise we apply only the new method. The solid line is the WDLF constructed from all stars regardless of the population. It is overestimated by $\sim$0.2\,dex at all magnitudes. An overestimation is expected but a consistent overdensity of 0.2\,dex is not negligible. Due to the spatial density correction, the maximum volume of an object integrated over a disc profile has smaller volume than a halo profile. While objects are weighted by the inverse maximum volume, the apparently negligible contribution from the older populations would by amplified by density correction such that each contaminating object would have a volume underestimated by tens of percentage points. The discovery fraction for a disc population is always smaller than that of the halo, so this amplifies the 1/V$_{max}$ by another few tens of percentage points. The V/V$_{max}$ distributions are consistently at around 0.55 as a consequence of the halo objects contribution. The ratio of the volume element between an exponential disc density profile and a uniform halo increases with distance. Therefore, for a group of uniformly distributed objects being assigned to follow an exponential profile in the integrals, the resulting $\langle$V/V$_{max}\rangle$ is expected to be larger than 0.5. The dashed line is constructed with an addition of a fixed upper tangential velocity limit at 80\,km\,s$^{-1}$ which would eliminate most of the halo objects, but not the ones from the thick disc. This reduces the contamination down to roughly 0.1\,dex at all magnitudes, with a smaller effect on the $\langle$V/V$_{max}\rangle$ distribution. The dotted line is computed using only thin disc objects from that simulation for comparison. Everything agrees to within statistical uncertainties at all magnitudes in the thin disc-only analysis. Since the discovery fraction requires the kinematics of the populations, a good membership association is very important.

The aforementioned RH11 effective volume method in Section~\ref{sec:new_approach} was not dealt with in detail because of some fundamental differences in the method. That method can untangle the three components of the Galaxy. However, the binning of objects at an early stage means that it is not directly compatible with the framework in this work which, instead, bins objects in the final step. In dealing with observational data without the labels ``thin disc", ``thick disc" and ``halo" tagged on the objects, the best one can do is to perform the membership assignment statistically. The effective volume method would provide a quick and easy way to do the job. However, if one would like to retain as much information as the maximum volume family would normally allow, the best way to do it is to find the probability of each object belonging to each of the components based on their observed and intrinsic properties, e.g. tangential velocities metallicities, and/or distance from the Galactic Plane. Population membership assignment is beyond the scope of this paper but if it were to be done in a maximum likelihood approach, readers are reminded that distributions are usually approximated as Gaussian functions. In the case of tangential velocity, the PDF in fact follows a Schwarzschild distribution in which case a Gaussian distribution cannot approximate the tail at high velocity well. It is important to assess whether this would lead to a bias in assigning objects to higher velocity populations. However, if one were to split the tangential velocity into ($l, b$) components, the distributions of the component velocities should be better approximated by Gaussian functions. As for the physical position, both the current distance from the plane and the maximum distance in which an object can reach based on the current kinematics, Z$_{\text{max}}$, would be good parameters to test. Both suggested parameters have used the important piece of information that a halo object is much more likely to carry a larger velocity perpendicular to the plane than those of the discs.

\subsection{Different Galactic Models}
\begin{figure}
\begin{center}
\includegraphics[width=84mm]{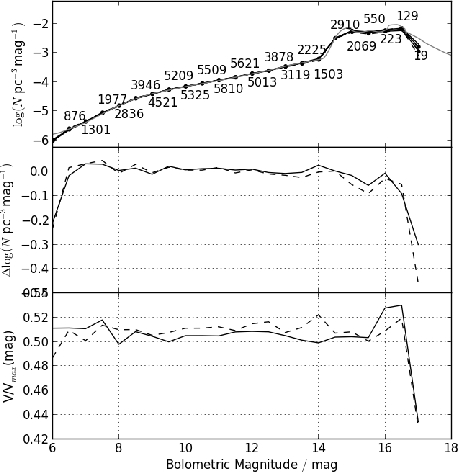}
\caption{
Top: WDLFs for the stellar halo with different input Galactic models. The solid line is the original halo WDLF identical to the one in Fig\,\ref{fig:wdlf_100_new_old}, the dashed line is the one that follows a simplified Besan\c{c}on model but then analyzed with the original set of kinematics. Middle and bottom panels are the same as before.}
\label{fig:different_models}
\end{center}
\end{figure}

The accuracy of all LF estimators is sensitive to the assumed population kinematics and density profiles. Therefore it is important to investigate the effects of applying a different kinematic model in the analysis of a population simulated with another model. To demonstrate the effect, we have adopted a simplified version of the Besan\c{c}on galaxy model \citep{2003A&A...409..523R,2004A&A...416..157R}. A halo population at 100 times the measured density with $(U, V, W)=(0.0,-220.0,0.0)$\,km\,s$^{-1}$ and velocity dispersion $(\sigma_{U}, \sigma_{V}, \sigma_{W})=(131.0, 106.0, 85.0)$\,km\,s$^{-1}$ was generated using the same recipe described in Section~\ref{sec:mcsimulation}. The halo population is then analyzed using the original set of kinematic properties. The same WDLF from Section~\ref{sec:stellar_halo_100} is overplotted for comparison in Fig.~\ref{fig:different_models}. The two WDLFs agree at all magnitudes and even the trends in underdensities at faint magnitudes are very similar. The differences in the $\langle$V/V$_{max}\rangle$ distribution at faint magnitudes can be explained by the same argument as in Section~\ref{sec:galaxy}. Due to the similar kinematic properties between the halo kinematic models applied in this work and the Besan\c{c}on model, the size of this effect is much smaller. From this simple experiment we conclude that small differences between the assumed and underlying kinematics results in little differences in both the derived WDLF and the $\langle$V/V$_{max}\rangle$ distribution.

\section{Conclusion}
We have presented an improved treatment of the proper motion and tangential velocity limits to arrive at an unbiased luminosity function estimator for stellar populations selected on the basis of both magnitude and proper motion. Our simulations have shown that the assumption of setting the tangential velocity of an object as its intrinsic property would be invalid when a sample is drawn from a survey that has  restricted proper motion limits and for populations where the tangential velocity distribution function varies over the sky. In order to select high quality proper motion objects when analysing real survey data the lower proper motion limit and tangential velocity limit are usually set at such a level that the velocity space is very restricted. Therefore, it is necessary to correlate the tangential velocity with the proper motion depending on the properties of each individual object. Two biases were identified from dissecting the generalized V$_{max}$ integral. The first one is an inconsistent density measurement of any non-uniform sample in which the density depends on the line of sight direction. The generalized method only considers the population spatial density and survey limits, so different densities would be measured at different line of sight directions as the distribution of proper motion and tangential velocity vary with line of sight. The other arises from the assumption that any given object would carry a constant tangential velocity. The choice of the upper proper motion limit has to be carefully determined from the properties of the survey. An arbitrarily large upper limit has negligible effects in the analysis as the survey volume is sensitive only to the lower proper motion limit. The discovery fraction is, however, sensitive to both upper and lower proper motion limits. From our simulations, it is found that surveys with upper proper motion limits smaller than $\sim$0.5" yr$^{-1}$ require more attention. This value seems small to the modern and future surveys~(e.g.~Pan--STARRS, Gaia, LSST etc.) but when a pairing is done with the early epoch Schmidt survey plate data for the bright objects for example the upper proper motion limits would shrink rapidly when the maximum epoch differences can be up to more than half a century (the pairing radius cannot be increased indefinitely when associating catalogues with very different depth and resolution without introducing many spurious source matches).

Population membership association has to be done carefully because with a wrong set of kinematics the derived set of discovery fractions would become completely meaningless. However, in the case that a population completely dominates the density budget, for example the thin disc objects in the solar neighbourhood, the wrong membership associations have negligible effects to the luminosity functions. Finally, we would like to remind the readers to consider the limitations of  $\langle$V/V$_{max}\rangle$ carefully: (i)\,a flat V/V$_{max}$ distribution with $\langle$V/V$_{max}\rangle \sim 0.5$ only indicates if the analysis is not biased towards distant or nearer objects, it gives no indication of the completeness of the sample. For a given flat distribution, one can conclude, at best, that the survey is equally sensitive to objects at any distance. However, (ii)\, given a complete sample, a flat distribution and $\langle$V/V$_{max}\rangle \sim 0.5$ are expected.

\subsection*{Application to other density estimators}
The treatment of detection limits due to proper motion selection have never been included in other density estimators. The modified Schmidt estimator is the only one that has been extended to incorporate such constraints. However, as detailed earlier, the treatment breaks down if we have restricted proper motion selection criteria. \citet{2006MNRAS.369.1654G} compared the properties of three different estimators: generalized Schmidt, Stepwise Maximum Likelihood and Choloniewski method. It is clear that the treatment of the proper motion limits were only applied to the Schmidt estimator by considering the tangential velocity as an intrinsic property of an object, while the other two considered the case of a magnitude limited sample. Their analysis may be affected by the modifications to the density estimator made in this work because the sample adopted in their work has assumed a fixed tangential velocity along line of sight. However, their work focused on the thin disc which is much less susceptible to the bias we identified, and by adopting a large upper proper motion limit in their data selection~($\mu \in \left[\,0.16,2.00\,\right]$), the effect should be tiny. Nonetheless we would like to remind readers to pay attention to the faint end of their analysis which is most affected if the bias is noticeable at all. It should be possible to generalize the discovery fraction (as a completeness correction arising from tangential velocity and/or proper motion limit) to other density estimators as the proper motion is now decoupled from the detection limit.

\subsection*{Small Scale Variations}
Modern digital surveys are probing larger volumes with multi-epoch measurements over large areas of sky (e.g.~SDSS, VST, Pan--STARRS, LSST and Gaia). Astrometric precision is, however, a strong function of the direction of the lines of sight due to varying observing qualities and imperfect camera fill factor arising from CCD chip gaps. In order to fully utilize the sample volume probed in any single or combined surveys, it is necessary to generalize the density estimators to cope with small scale fluctuations of the survey properties. Otherwise, the volume of the survey must be limited to the shallowest part of the sky. One possible solution is to pixelise the sky into equal area pixels using HEALPix\footnote{http://healpix.jpl.nasa.gov/}. The photometric and astrometric precision of each pixel are then limited to the worst observing condition within the pixel.

\subsection*{Galactic model}
The current Schwarzschild distribution assumes a Cartesian system centred at the Sun. This approximation is good for surveys that probe only small distances~(a few hundred parsecs). However, in the future surveys, most notably the Gaia and LSST, where the lower proper motion limits at the detection limits would be as low as 0.2\,mas\,yr$^{-1}$~(G$_{Gaia}\sim20$\,mag), and 1\,mas\,yr$^{-1}$~(r$_{LSST}\sim24$\,mag) respectively, an object with a tangential velocity of 20\,km\,s$^{-1}$ could be detected with 5$\sigma$ confidence at $\sim$17\,kpc and $\sim$3.5\,kpc respectively. A WD with T$_{\text{eff}} \sim$ 3000\,K and M$_{bol} \approx$16\,mag, which translate to G$_{\text{Gaia}}\sim$15.8\,mag and g$_{\text{LSST}}\sim$17.3\,mag at 10\,pc, would be detectable at $\sim$1-2\,kpc in both cases. The overdensity at the spiral arms is not accounted for in any work to date because the current surveys have not reached such distances where the nearest spiral arm lies. However, when the faintest objects can be detected at a distance of over a thousand parsecs, the brighter objects would lie easily inside the spiral arm region which is $\sim$1\,kpc from the Sun. In such a situation it is necessary to employ a more detailed density profile to account for the varying density in not just the vertical direction but also in the planar directions. With the decoupling of the proper motion limit from the photometric limits, the incorporation of a more sophisticated model would simply be a translation of the Galactic density and velocity profiles from the galactocentric frame to the geocentric frame. The key difference is that a Schwarzschild distribution function gives a constant tangential velocity distribution function as a function of the line of sight\,(ie. $\chi = \chi(\alpha,\delta)$), but in the adoption of a complex model, the density and velocity distribution functions have to be found as functions of both distance and the direction of line of sight such that $\chi = \chi(\alpha,\delta,D)$.

\section*{Acknowledgements}
The Pan--STARRS1 Surveys\,(PS1) have been made possible through contributions of the Institute for Astronomy, the University of Hawaii, the Pan--STARRS Project Office, the Max-Planck Society and its participating institutes, the Max Planck Institute for Astronomy, Heidelberg and the Max Planck Institute for Extraterrestrial Physics, Garching, The Johns Hopkins University, Durham University, the University of Edinburgh, Queen's University Belfast, the Harvard-Smithsonian Center for Astrophysics, the Las Cumbres Observatory Global Telescope Network Incorporated, the National Central University of Taiwan, the Space Telescope Science Institute, the National Aeronautics and Space Administration under Grant No. NNX08AR22G issued through the Planetary Science Division of the NASA Science Mission Directorate, the National Science Foundation under Grant No. AST-1238877, the University of Maryland, and Eotvos Lorand University\,(ELTE). ML acknowledges financial support from a UK STFC PhD studentship.

We thank the PS1 Builders and PS1 operations staff for construction and operation of the PS1 system and access to the data products provided. We also thank the referee for pointing out the use of confusing terminology and diagrams.
\bibliography{2015_vmax}

\label{lastpage}

\end{document}